# Science of Cyber Security as a System of Models and Problems

Alexander Kott, US Army Research Laboratory, Adelphi, MD

## Abstract

Terms like "Science of Cyber" or "Cyber Science" have been appearing in literature with growing frequency, and influential organizations initiated research initiatives toward developing such a science even though it is not clearly defined. We propose to define the domain of the science of cyber security by noting the most salient artifact within cyber security -- malicious software -- and defining the domain as comprised of phenomena that involve malicious software (as well as legitimate software and protocols used maliciously) used to compel a computing device or a network of computing devices to perform actions desired by the perpetrator of malicious software (the attacker) and generally contrary to the intent (the policy) of the legitimate owner or operator (the defender) of the computing device(s). We further define the science of cyber security as the study of relations – preferably expressed as theoretically-grounded models -- between attributes, structures and dynamics of: violations of cyber security policy; the network of computing devices under attack; the defenders' tools and techniques; and the attackers' tools and techniques where malicious software plays the central role. We offer a simple formalism of these key objects within cyber science and systematically derive a classification of primary problem classes within cyber science.

## Introduction

Few things are more suspect than a claim of the birth of a new science. Yet, in the last few years, terms like "Science of Cyber," or "Science of Cyber Security," or "Cyber Science" have been appearing in use with growing frequency. For example, the US Department of Defense defined "Cyber Science" as a high priority for its science and technology investments (Lemnios 2011), and the National Security Agency has been exploring the nature of the "science of cybersecurity" in its publications, e.g., (Longstaff 2012). This interest in science of cyber is motivated by the recognition that development of cyber technologies is handicapped by the lack of scientific understanding of the cyber phenomena, particularly the fundamental laws, theories, and theoretically-grounded and empirically validated models (JASON 2010). Lack of such fundamental knowledge – and its importance – has been highlighted by the US President's National Science and

Technology Council (NSTC 2011) using the term "cybersecurity science."

Still, even for those in the cyber security community who agree with the need for science of cyber -- whether it merits an exalted title of a new science or should be seen merely as a distinct field of research within one or more of established sciences -- the exact nature of the new science, its scope and boundaries remain rather unclear.

This chapter offers an approach to describing this scope in a semi-formal fashion, with special attention to identifying and characterizing the classes of problems that the science of cyber should address. In effect, we will map out the landscape of the science of cyber as a coherent classification of its characteristic problems. Examples of current research -- mainly taken from the portfolio of the United States Army Research Laboratory where the author works -- will illustrate selected classes of problems within this landscape.

## Defining the Science of Cyber Security

A research field -- whether or not we declare it a distinct new science -- should be characterized from at least two perspectives. First is the domain or objects of study, i.e., the classes of entities and phenomena that are being studied in this research field. Second is the set of characteristic problems, the types of questions that are asked about the objects of study. Related examples of attempts to define a field of research include (Willis 20000) and (Bostrom 2007).

To define the domain of the science of cyber security, let's focus on the most salient artifact within cyber security -- malicious software. This leads us to the following definition: the domain of science of cyber security is comprised of phenomena that involve malicious software (as well as legitimate software and protocols used maliciously) used to compel a computing device or a network of computing devices to perform actions desired by the perpetrator of malicious software (the attacker) and generally contrary to the intent (the policy) of the legitimate owner or operator (the defender) of the computing device(s). In other words, the objects of research in cyber security are:

- Attacker $A$ along with the attacker's tools (especially malware) and techniques $T_a$
- Defender $D$ along with the defender's defensive tools and techniques $T_d$, and operational assets, networks and systems $N_d$
- Policy $P$, a set of defender's assertions or requirements about what event should and should not happen. To simplify, we may focus on cyber incidents $I$: events that should not happen.

Note that this definition of relevant domain helps to answer common questions about the relations between cyber security and established fields like electronic warfare and cryptology. Neither electronic warfare nor cryptology focus on malware and processes pertaining to malware as the primary objects of study.

The second aspect of the definition is the types of questions that researchers ask about the objects of study. Given the objects of cyber security we proposed above, the primary questions revolve around the relations between $T_a$, $T_d$, $N_d$, and $I$ (somewhat similar perspective is suggested in (Schneider 2012) and in (Bau and Mitchell 2011). A shorthand for the totality of such relations might be stated as

$$(I, T_d, N_d, T_a) = 0 \qquad (1)$$

This equation does not mean I expect to see a fundamental equation of this form. It is merely a shorthand that reflects our expectation that cyber incidents (i.e., violations of cyber security policy) depend on attributes, structures and dynamics of the network of computing devices under attack, and the tools and techniques of defenders and attackers.

Let us now summarize what we discussed so far in the following definition. The science of cyber security is the study of relations between attributes, structures and dynamics of: violations of cyber security policy; the network of computing devices under attack; the defenders' tools and techniques; and the attackers' tools and techniques where malicious software plays the central role.

A study of relations between properties of the study's objects finds its most tangible manifestation in models and theories. The central role of models in science is well recognized; it can be argued that a science is a collection of models (Frigg 2012), or that a scientific theory is a family of models or a generalized schema for models (Cartwright 1983, Suppes 2002). From this perspective, we can restate our definition of science of cyber security as follows. The science of cyber security develops a coherent family of models of relations between attributes, structures and dynamics of: violations of cyber security policy; the network of computing devices under attack; the defenders' tools and techniques; and the attackers' tools and techniques where malicious software plays the central role. Such models

1. are expressed in an appropriate rigorous formalism;
2. explicitly specify assumptions, simplifications and constraints;
3. involve characteristics of threats, defensive mechanisms and the defended network;
4. are at least partly theoretically grounded;
5. yield experimentally testable predictions of characteristics of security violations.

There is a close correspondence between a class of problems and the models that help solve the problem. The ensuing sections of this chapter look at specific classes of problems of cyber security and the corresponding classes of models. We find that Eq. 1 provides a convenient basis for deriving an exhaustive set of such problems and models in a systematic fashion.

## Development of Intrusion Detection Tools

Intrusion detection is one of the most common subjects of research literature generally recognized as falling into the realm of cyber security. Much of research in intrusion detection focuses on proposing novel algorithms and architectures of intrusion detection tools. A related topic is characterization of efficacy of such tools, e.g., the rate of detecting true intrusions or the false alert rate of a proposed tool or algorithm in comparison with prior art.

To generalize, the problem addressed by this literature is to find (also, to derive or synthesize) an algorithmic process , or technique, or architecture of defensive tool that detects certain types of malicious activities, with given assumptions (often implicit) about the nature of computing devices and network being attacked, about the defensive policies (e.g., a requirement for rapid and complete identification of intrusions or information exfiltration, with high probability of success), and about the general intent and approaches of the attacker. More formally, in this problem we seek to derive $T_d$ from $N_d$, $T_a$, and $I$, i.e.,

$$N_d, T_a, I \rightarrow T_d \qquad (2)$$

Recall that $T_d$ refers to a general description of defenders' tools and techniques, that may include an algorithmic process or rules of an intrusion detection tool, as well as architecture of an IDS or IPS, and attributes of an IDS such as its detection rate. In other words, Eq. 2 is shorthand for a broad class of problems. Also note that Eq. 2 is derived from Eq. 1 by focusing on one of the terms on the left hand side of Eq. 1.

To illustrate the breadth of issues included this class of problems, let's consider an example – a research effort conducted at the US Army Research Laboratory that seeks architectures and approaches to detection of intrusions in a wireless mobile network (Ge et al. 2012). In this research, we make an assumption that the intrusions are of a sophisticated nature and are unlikely to be detected by a signature-matching or anomaly-based algorithm. Instead, it requires a comprehensive analysis and correlation of information obtained from multiple devices operating on the network, performed by a comprehensive collection of diverse tools and by an insightful human analysis.

One architectural approach to meeting such requirements would comprise multiple software agents deployed on all or most of the computing devices of the wireless network; the agents would send their observations of the network traffic and of host-based activities to a central analysis facility; and the central analysis facility would perform a comprehensive processing and correlation of this information, with participation of a competent human analyst.

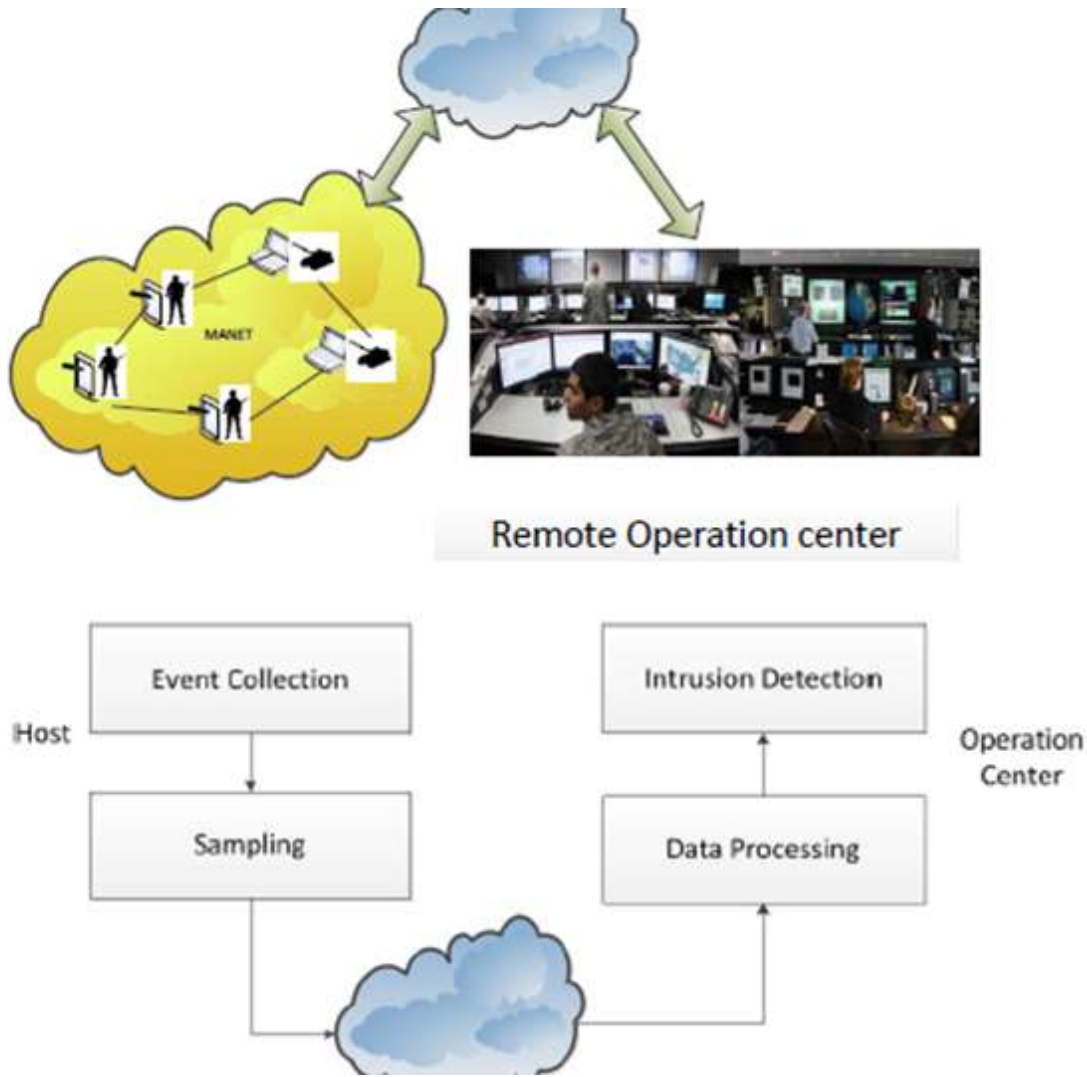

Fig.1. Local agents on the hosts of the mobile network collect and sample information about host-based and network events; this information is aggregated and transmitted to the operation center where comprehensive analysis and detection are performed. Adapted from (Ge et. al. 2012) with permission.

Such an approach raises a number of complex research issues. For example, because the bandwidth of the wireless network is limited by a number of factors, it is desirable to use appropriate sampling and in-network aggregation and pre-processing of information produced by the local software agents before transmitting all this information to the central facility. Techniques to determine appropriate locations for such intermediate aggregation and processing are needed. Also needed are algorithms for performing aggregation and pre-processing that maximize the likelihood of preserving the critical information indicating an intrusion.  We also wish to have means to characterize the resulting detection accuracy in this bandwidth-restricted, mobile environment.

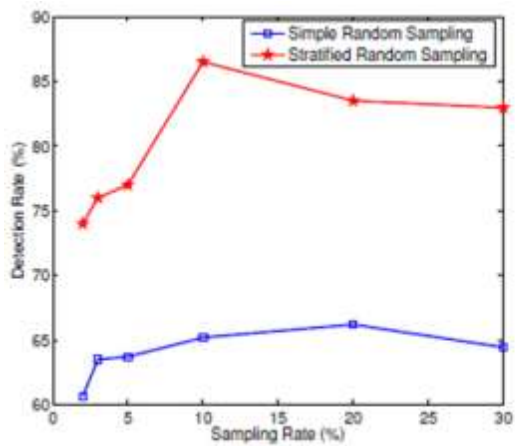 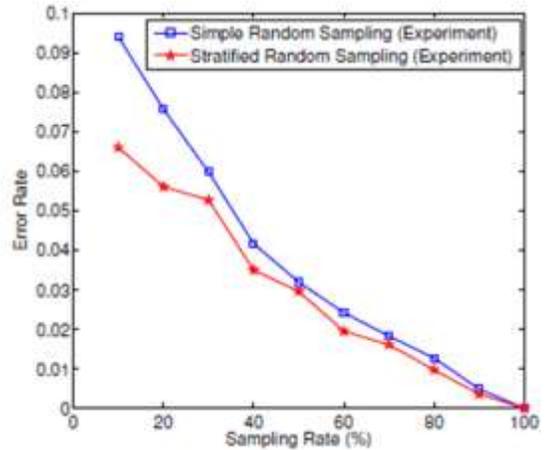

Fig. 2. Experiments suggest that detection rates and error rate of detection strongly depend on the traffic sampling ratio as well as the specific strategy of sampling. Adapted from (Ge et al 2012) with permission.

Eq.2 captures key elements of this class of problems. E.g., $T_d$ in Eq.2 is the abstraction of this defensive tool's structure (e.g., locations of interim processing points), behavior (e.g., algorithms for pre-processing), and attributes (e.g., detection rate). Designers of such a tool would benefit from a model that predicts the efficacy of the intrusion detection process as a function of architectural decisions, properties of the algorithms and properties of the anticipated attacker's tools and techniques.

## Cyber Maneuver and Moving Target Defense

Cyber maneuver refers to the process of actively changing the defended network -- its topology, allocation of functions and properties (Jajodia et al. 2011). Such changes can be useful for several reasons. Continuous changes help to confuse the attacker and to reduce the attacker's ability to conduct effective reconnaissance of the network in preparation for an attack. This use of cyber maneuver is also called moving target defense. Other types of cyber maneuver could be used to minimize effects of an ongoing attack, to control damage, or to restore the network's operations after an attack.

Specific approaches to cyber maneuver and moving target defense, such as randomization and enumeration are discussed in (Jajodia et al. 2011, Jajodia et al. 2013). Randomization can take multiple forms: memory address space layout, (e.g., (Bojinov 2011]); instruction set (Barrantes 2005, Boyd 2010); compiler-generated software diversity; encryption; network address and layout; service locations, traffic patterns; task replication and breakdown across cores or machines; access policies; virtualization; obfuscation of OS types and services; randomized and multi-path routing, and others. Moving Target Defense has been identified as one of four strategic thrusts in the strategic plan for cyber security developed by the National Science and Technology Council (NSTC 2011).

Depending on its purpose, the cyber maneuver involves a large number of changes to the network executed by the network's defenders rapidly and potentially continuously over a long period of time. The defender's challenge is to plan this complex sequence of actions and to control its execution in such a way that the maneuver achieves its goals without destabilizing the network or confusing its users; a challenge that has been extensively explored but not necessarily solved in AI planning literature, e.g., (Sadeh and Kott 1996; Kott 2006).

Until now, we used $T_d$ to denote the totality of attributes, structure and dynamics of the defender's tools and techniques. Let's introduce additional notation, where $ST_d$ is the structure of the defensive tools, and $BT_d(t)$ is the defender's actions. Then, referring to Eq. 1 and focusing on $BT_d$ -- the sub-element of $T_d$, the class of problems related to synthesis and control of defenders course of action can be described as

$$N_d, T_a, I \rightarrow BT_d(t) \qquad (3)$$

An example of problem in this class is to design a technique of cyber maneuver in a mobile ad hoc spread-spectrum network where some of the nodes are compromised via a cyber attack and become adversary-controlled jammers of the network's communications. One approach is to execute a cyber maneuver using spread-spectrum keys as maneuver keys (Torrieri, Zhu and Jajodia 2013). Such keys supplement the higher-level network cryptographic keys and provide the means to resist and respond to external and insider attacks. The approach also includes components for attack detection, identification of compromised nodes, and group rekeying that excludes compromised nodes.

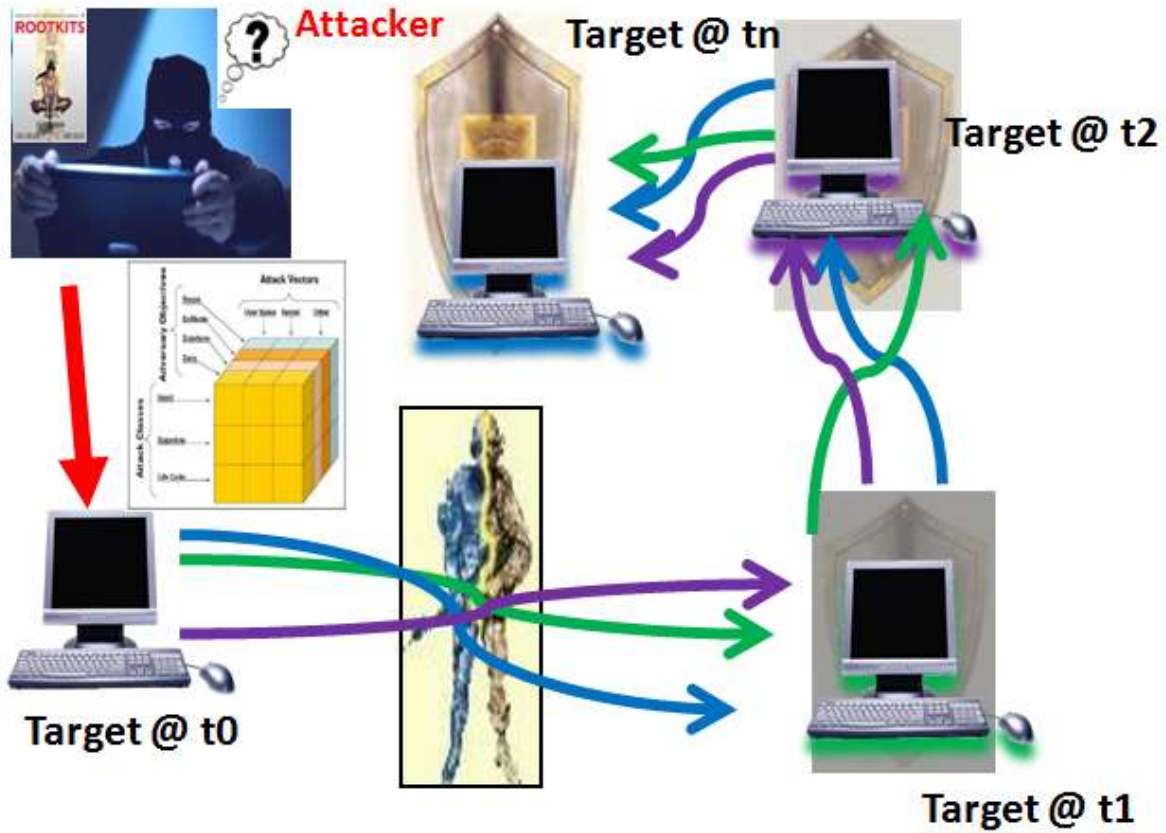

Fig. 3. In moving target defense, the network continually changes its attributes visible to the attacker, in order to minimize the attacker's opportunities for planning and executing an effective attack.

Eq. 3 captures the key features of the problem: we wish to derive the plan of cyber maneuver $BT_d(t)$ from known or estimated changes in properties of our network, properties of anticipated or actually observed attacks $T_a$, and the objective of minimizing security violations $I$. Planning and execution of a cyber maneuver would benefit from models that predict relevant properties of the maneuver, such as its convergence to a desired end state, stability, or reduction of observability to the attacker.

## Assessment of Network's Vulnerabilities and Risks

Monitoring and assessment of vulnerabilities and risks is an important part of cyber security strategy pursued by the US Government (Dempsey et al. 2011) This involves continuous collection of data through automated feeds including network traffic information as well as host information from host-based agents: vulnerability information and patch status about hosts on the network; scan results from tools like Nessus; TCP netflow data; DNS trees, etc. These data undergo automated analysis in order to assess the risks. The assessment may include flagging especially egregious vulnerabilities and exposures, or computing metrics that provide an overall characterization of the network's risk level. In current practice, risk metrics are often simple sums or counts of vulnerabilities and missing patches.

There are important benefits in automated quantification of risk, i.e., of assigning risk scores or other numerical measures to the network as w hole, its subsets and even individual assets (Kott and Arnold 2013; Gil 2014). This opens doors to true risk management decision-making, potentially highly rigorous and insightful. Employees at multiple levels – from senior leaders to system administrators – will be aware of continually updated risk distribution over the network components, and will use this awareness to prioritize application of resources to most effective remedial actions. Quantification of risks can also contribute to rapid, automated or semi-automated implementation of remediation plans.

However, existing risk scoring algorithms remain limited to ad hoc heuristics such as simple sums of vulnerability scores or counts of things like missing patches or open ports, etc.  Weaknesses and potentially misleading nature of such metrics have been pointed out by a number of specialists, e.g., (Jensen 2009; Bartol 2009). For example, the individual vulnerability scores are dangerously reliant on subjective, human, qualitative input, potentially inaccurate and expensive to obtain. Further, the total number of vulnerabilities may matters far less than how vulnerabilities are distributed over hosts, or over time. Similarly, neither topology of the network nor the roles and dynamics of inter-host interactions are considered by simple sums of vulnerabilities or missing patches. In general, there is a pronounced lack of rigorous theory and models of how various factors might combine into quantitative characterization of true risks, although there are initial efforts, such as (Lippman 2012) to formulate scientifically rigorous methods of calculating risks.

Returning to Eq. 1 and specializing the problem to one of finding $N_d$, we obtain

$$I, T_d, T_a \rightarrow N_d \qquad (4)$$

Recall that $N_d$ refers to the totality of the defender's network structure, behavior and properties. Therefore, Eq. 3 refers to a broad range of problems including those of synthesizing the design, the operational plans and the overall properties of the network we are to defend. Vulnerabilities, risk, robustness, resiliency and controllability of a network are all examples of the network's properties, and Eq. 3 captures the problem of modeling and computing such properties.

An example of research on the problem of developing models of properties of robustness, resilience, network control effectiveness, and collaboration in networks is (Cam 2012). The author explores approaches to characterizing the relative criticality of cyber assets by taking into account risk assessment (e.g., threats, vulnerabilities), multiple attributes (e.g., resilience, control, and influence), network connectivity and controllability among collaborative cyber assets in networks. In particular, the interactions between nodes of the network must be considered in assessing how vulnerable they are and what mutual defense mechanisms are available.

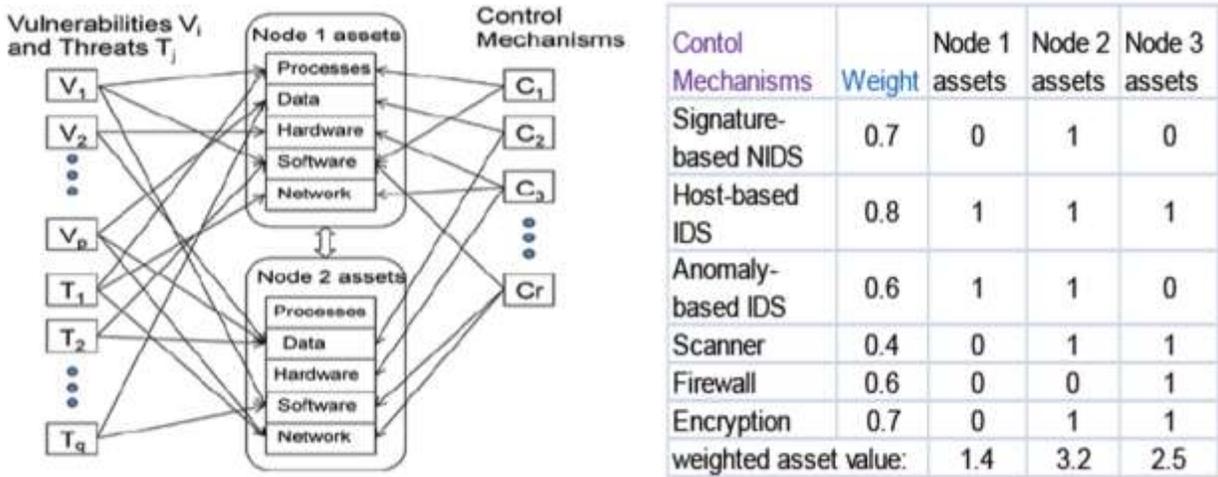

Fig. 4. Risk assessment of a network must take into account complex interactions between nodes of the network, particularly the interactions between their vulnerabilities as well as opportunities for mutual defense. Adapted from (Hasan 2012 with permission.

## Attack Detection and Prediction

Detection of malicious activities on networks is among the oldest and most common problems in cyber security (Anderson 1980). A broad subset of such problems is often called intrusion detection. Approaches to intrusion detection are usually divided into two classes, signature-based approaches and anomaly-based approach, both with their significant challenges (Axelsson 2000, Patcha and Park 2007). In Eq. 3, the term $I$ refers to malicious activities or intrusions, including structures, behaviors and properties of such activities. Therefore, the process of determining whether a malicious activity is present and its timing, location and characteristics, are reflected in the following expression:

$$T_d, N_d, T_a \rightarrow I \qquad (5)$$

The broad class of problems captured by Eq. 5 includes the problem of deriving key properties of a malicious activity, including the very fact of an existence of such an activity, from the available information about the tools and techniques of the attacker $T_a$ (e.g., the estimated degree of sophistication and the nature of past attempted attacks of the likely threats), tools and techniques of the defender $T_d$ (e.g., locations and capabilities of the firewalls and intrusion-prevention systems), and the observed events on the defender's network $N_d$ (e.g., the alerts received from host based agents or network based intrusion detection systems).

Among the formidable challenges of the detection problem is the fact that human analysts and their cognitive processes are critical components within the modern practices of intrusion detection. However, the human factors and their properties in cyber security have been inadequately studied and are poorly understood (McNeese 2012, Boyce et al. 2011).

Unlike the detection problem that focuses on identifying and characterizing malicious activities that have already happened or at least have been initiated, i.e., *I(t)* for $t<t_{now}$, the prediction problem seeks to characterize malicious activities that are to occur in the future, i.e., *I(t)* for $t>t_{now}$. The extent of research efforts and the resulting progress has been far less substantial in prediction than in detection, although related research has occurred in related adversarial domains, e.g., (Ownby and Kott 2006). Theoretically grounded models that predict characteristics of malicious activities *I* – including the property of detectability of the activity -- as a function of $T_a$, $T_d$, $N_a$, would be major contributors into advancing this area of research.

An example of research on identifying and characterizing probable malicious activities, with a predictive element as well, is (Harang and Glodek 2012), where the focus is on fraudulent use of security tokens for unauthorized access to network resources. Authors explore approaches to detecting such fraudulent access instances through a network-based intrusion detection system that uses a parsimonious set of information. Specifically, they present an anomaly detection system based upon IP addresses, a mapping of geographic location as inferred from IP address, and usage timestamps. The anomaly detector is capable of identifying fraudulent token usage with as little as a single instance of fraudulent usage while overcoming the often significant limitations in geographic IP address mappings. This research finds significant advantages in a novel unsupervised learning approach to authenticating fraudulent access attempts via time/distance clustering on sparse data.

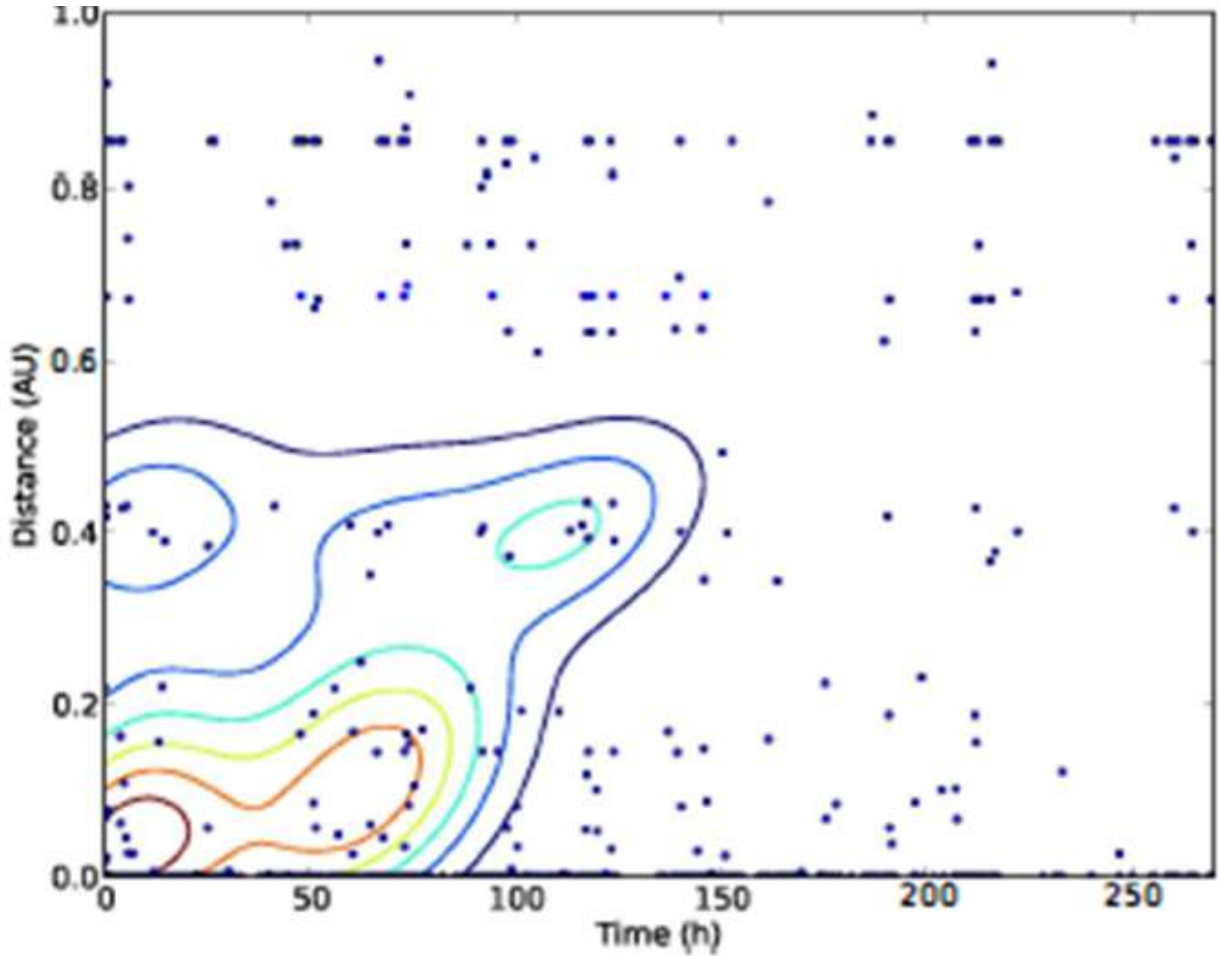

Fig. 5. There exist multiple patterns of time-distance pairs of legitimate users' subsequent log-ins. This figure depicts the pattern of single location users combined with typical commuters that log-in from more than one location. Adapted from (Harang and Glodek 2012) with permission.

## Threat Analysis and Cyber Wargaming

Returning once more to Eq. 1, consider the class of problems where the tools and techniques of the attacker Ta are of primary interest:

$$I, T_d, N_d, \rightarrow T_a \qquad (6)$$

Within this class of problems we see for example the problem of deriving structure, behavior and properties of malware from the examples of the malicious code or from partial observations of its malicious activities. Reverse engineering and malware analysis, including methods of detecting malware by observing a code's structure and characteristics, fall into this class of problems.

A special subclass of problems occurs when we focus on anticipating the behavior of attacker over time as a function of defender's behavior:

$$I(t), T_d(t), N_d, \rightarrow T_a(t) \qquad (6a)$$

In this problem, game considerations are important -- both the defender's and the attacker's actions are at least partially strategic and depend on their assumptions and anticipations of each other's actions. Topics like adversarial analysis and reasoning, wargaming, anticipation of threat actions, and course of action development fall into this subclass of problems.

## Summary of the Cyber Science Problem Landscape

We now summarize the classification of major problem groups in cyber security. All of these derive from Eq. 1. For each subclass, an example of a common problem in cyber security research and practice is added, for the sake of illustration.

$T_d, T_a, I \rightarrow N_d$

    $T_d, T_a, I \rightarrow SN_d(t)$ – e.g., synthesis of network's structure

    $T_d, T_a, I \rightarrow BN_d(t)$ – e.g., planning and anticipation of network's behavior

    $T_d, T_a, I \rightarrow PN_d(t)$ – e.g., assessing and anticipating network's security properties

$N_d, T_a, I \rightarrow T_d$

    $N_d, T_a, I \rightarrow ST_d(t)$ – e.g., design of defensive tools, algorithms

    $N_d, T_a, I \rightarrow BT_d(t)$ – e.g., planning and control of defender's course of action

    $N_d, T_a, I \rightarrow PT_d(t)$ – e.g., assessing and anticipating the efficacy of defense

$T_d, N_d, I \rightarrow T_a$

    $T_d, N_d, I \rightarrow ST_a(t)$ – e.g., identification of structure of attacker's code or infrastructure

    $T_d, N_d, I \rightarrow BT_a(t)$ – e.g., discovery, anticipation and wargaming of attacker' actions

    $T_d, N_d, I \rightarrow PT_a(t)$ – e.g., anticipating the efficacy of attacker's actions

$T_d, N_d, T_a \rightarrow I$

$T_d, N_d, T_a \rightarrow I(t), t<t_{now}$ – e.g., detection of intrusions that have occured

$T_d, N_d, T_a \rightarrow I(t), t>t_{now}$ – e.g., anticipation of intrusions that will occur

## Conclusions

As a research field, the emerging science of cyber security can be defined as the search for a coherent family of models of relations between attributes, structures and dynamics of: violations of cyber security policy; the network of computing devices under attack; the defenders' tools and techniques; and the attackers' tools and techniques where malicious software plays the central role. As cyber science matures, it will see emergence of models that should: (a) be expressed in an appropriate rigorous formalism; (b) explicitly specify assumptions, simplifications and constraints; (c) involve characteristics of threats, defensive mechanisms and the defended network; (c) be at least partly theoretically grounded; and (d) yield experimentally testable predictions of characteristics of security violations. Such models are motivated by key problems in cyber security. We propose and systematically derive a classification of key problems in cyber security, and illustrate with examples of current research.